\documentclass[aps,physrev,onecolumn,superscriptaddress,
longbibliography,floatfix]{revtex4-2}

\usepackage{amsmath,amssymb,graphicx,bm}
\usepackage[colorlinks=true,
citecolor=blue,
linkcolor=blue,
urlcolor=blue]{hyperref}
\usepackage[caption=false]{subfig}
\usepackage{placeins}

\DeclareMathOperator{\sgn}{sgn}

\begin{document}

\title{Lossless propagation of gain-compensated graphene plasmons}

\author{Andrianos Sygrimis}
\affiliation{Department of Physics, University of Crete, Heraklion 71003,
Greece}

\author{Giorgos P. Tsironis}
\affiliation{Department of Physics, University of Crete, Heraklion 71003,
Greece}
\affiliation{School of Engineering and Applied Sciences, Harvard University,
Cambridge, MA, USA}

\begin{abstract}
Graphene plasmons offer strong electromagnetic confinement and electrical
tunability, but dissipation in the graphene sheet limits their propagation.
Here, we investigate how optical gain in the surrounding dielectrics can
compensate these losses in single- and double-layer graphene. Starting from the
full electromagnetic boundary conditions and a finite-temperature graphene
conductivity, we derive the compensation condition for a single sheet and the
separate optical and acoustic dispersion relations for two coupled sheets. We
show that the required gain depends both on material loss and on where the gain
is placed. When gain is present throughout the dielectric environment, the two
double-layer modes have nearly identical compensation thresholds. When it is
confined to the spacer or the outer claddings, their thresholds separate because
the modes overlap differently with each region. Full-wave finite-element
simulations reproduce the predicted branches and field symmetries and show a
strong suppression of attenuation together with a substantial increase in
propagation length. The analysis also identifies the range over which a local
conductivity provides a controlled description. These results show how the
spatial distribution of gain can be used to control the compensation thresholds
of the optical and acoustic graphene plasmons.
\end{abstract}

\maketitle

\section{Introduction}

Surface plasmon polaritons (SPPs) are electromagnetic surface waves coupled to
collective charge oscillations at a conductor--dielectric interface
\cite{ritchie1957plasma,economou1969surface,pitarke2006theory}. Their
subwavelength confinement and enhanced near fields support nanophotonic routing,
sensing, and strong light--matter coupling \cite{barnes2003surface,
ozbay2006plasmonics,
schuller2010plasmonics,
gramotnev2010plasmonics,
maier2007plasmonics,
Han_2013}. Full-wave methods such as the finite element method (FEM) provide the
dispersion and field profiles of these modes in structured geometries
\cite{smajic2009comparison,jin2014finite}.

Graphene provides a two-dimensional (2D) conducting sheet whose plasmon
frequency and confinement can be tuned by carrier density
\cite{castro2009electronic,
peres2010colloquium,
jablan2009plasmonics,
koppens2011graphene,
jablan2013plasmons,
gonccalves2016introduction}. Gate-tunable graphene plasmons have been observed
directly by near-field imaging
\cite{fei2012gate,chen2012optical,woessner2015highly}. Their strong confinement,
however, is accompanied by dissipation through the real part of the graphene
conductivity, and additional microscopic channels can contribute to the
linewidth \cite{grigorenko2012graphene,principi2014plasmon}.

One route to longer propagation is to embed graphene in an active dielectric, so
that power supplied by the dielectric compensates the sheet loss
\cite{yu2008gain,
athanasopoulos2013enhanced,
bergman2003spaser,
noginov2009spaser,
oulton2009plasmon}. Gain compensation for an isolated graphene sheet and
graphene ribbons has already been analyzed, including analytical expressions in
the Drude and electrostatic limits
\cite{menendez2016graphene,zhernovnykova2019surface}; related theoretical
methods for graphene and active metamaterials are reviewed in
Ref.~\cite{mattheakis2017chapter}. We use SLG as an exact full-EM reference and
numerical benchmark, and then consider the coupled optical and acoustic modes of
DLG. Their separate dispersion equations reveal how placing gain in the spacer
or in the outer dielectrics changes the compensation threshold through the
distinct modal overlaps of the two branches.

Throughout the paper, fields vary as $\exp(iqx-i\omega t)$ and
\begin{equation}
 q=q'+iq'',\qquad \sigma=\sigma'+i\sigma'' .
 \label{eq:conventions}
\end{equation}
Thus $q''>0$ denotes attenuation and a dielectric
$\varepsilon=\varepsilon'+i\varepsilon''$ is active when $\varepsilon''<0$. For
a passive mode the intensity propagation length is $L=1/(2q'')$. The phrase
lossless propagation refers here to $q''=0$ in the local, linear material model;
gain saturation and additional microscopic loss channels are outside its scope.

\section{Graphene conductivity}

The sheet conductivity is evaluated with the finite-temperature local Kubo
model,
\begin{equation}
 \sigma(\omega)=\sigma_{\mathrm{intra}}(\omega)+\sigma_{\mathrm{inter}}(\omega),
 \label{eq:conductivity}
\end{equation}
where \cite{falkovsky2008optical,gusynin2007ac,peres2010colloquium}
\begin{equation}
 \sigma_{\mathrm{intra}}=
 \frac{2e^2k_BT}{\pi\hbar^2}
 \ln\!\left[2\cosh\!\left(\frac{E_F}{2k_BT}\right)\right]
 \frac{i}{\omega+i/\tau}
 \label{eq:intraband}
\end{equation}
and
\begin{align}
 \sigma_{\mathrm{inter}}=\frac{e^2}{4\hbar}\biggl[
 G\!\left(\frac{\hbar\omega}{2}\right)
 +\frac{4i\hbar\omega}{\pi}
 \int_0^\infty dE
 \times\frac{G(E)-G(\hbar\omega/2)}{\hbar^2\omega^2-4E^2}
 \biggr],
 \label{eq:interband}
\end{align}
with
\begin{equation}
 G(E)=\frac{\sinh(E/k_BT)}{
 \cosh(E_F/k_BT)+\cosh(E/k_BT)}.
\end{equation}
All analytical and numerical results use $E_F=0.20~\mathrm{eV}$,
$\tau=100~\mathrm{fs}$, $T=300~\mathrm{K}$, and $v_F=10^6~\mathrm{m\,s^{-1}}$.
The same intraband-plus-interband conductivity is used in the analytical
calculation and in COMSOL.

\section{Single-layer graphene}

For a transverse-magnetic (TM) mode at an interface between dielectrics
$\varepsilon_1$ and $\varepsilon_2$, define
\begin{equation}
 \kappa_j=\sqrt{q^2-\varepsilon_jk_0^2},\qquad k_0=\frac{\omega}{c},
\end{equation}
with the square-root branch chosen so that $\operatorname{Re}\kappa_j>0$. The
sheet-current boundary condition gives
\cite{hanson2008dyadic,jablan2009plasmonics}
\begin{equation}
 \frac{\varepsilon_1}{\kappa_1}+\frac{\varepsilon_2}{\kappa_2}
 +i\frac{\sigma}{\omega\varepsilon_0}=0.
 \label{eq:slg_general}
\end{equation}
For the symmetric case $\varepsilon_1=\varepsilon_2=\varepsilon_d$,
Eq.~\eqref{eq:slg_general} can be solved exactly. Introducing the dimensionless
conductivity $\bar\sigma=Z_0\sigma=\sigma/(\varepsilon_0c)$ and the effective
index $n=q/k_0$ gives
\begin{align}
 n^2&=\varepsilon_d-\frac{4\varepsilon_d^2}{\bar\sigma^2},\nonumber\\
 \frac{q}{k_0}=n&=
 \sqrt{\varepsilon_d-
 \frac{4\omega^2\varepsilon_0^2\varepsilon_d^2}{k_0^2\sigma^2}}.
 \label{eq:slg_index}
\end{align}
The second expression is the same result written in terms of $\sigma$, using
$\bar{\sigma}=\sigma/(\varepsilon_0c)$ and $k_0=\omega/c$.

To make the loss-compensation condition explicit, write
\begin{equation}
 \varepsilon_d=\varepsilon_d'+i\varepsilon_d'',\qquad
 \bar{\sigma}=\bar{\sigma}'+i\bar{\sigma}'',
\end{equation}
and denote the radicand in Eq.~\eqref{eq:slg_index} by $R=r+is$. Direct
expansion gives
\begin{align}
 r={}&\varepsilon_d'
 -\frac{4}{D}\bigl\{
 [(\varepsilon_d')^2-(\varepsilon_d'')^2]
 [(\bar{\sigma}')^2-(\bar{\sigma}'')^2]\nonumber\\
 &\hspace{4.2em}
 +4\varepsilon_d'\varepsilon_d''
 \bar{\sigma}'\bar{\sigma}''\bigr\},
 \label{eq:r}\\
 s={}&\varepsilon_d''
 -\frac{8}{D}\bigl\{
 \varepsilon_d'\varepsilon_d''
 [(\bar{\sigma}')^2-(\bar{\sigma}'')^2]\nonumber\\
 &\hspace{4.2em}
 -\bar{\sigma}'\bar{\sigma}''
 [(\varepsilon_d')^2-(\varepsilon_d'')^2]\bigr\},
 \label{eq:s}\\
 D={}&[(\bar{\sigma}')^2+(\bar{\sigma}'')^2]^2.
\end{align}
The corresponding complex square root is
\begin{equation}
 n=\sqrt{\frac{\sqrt{r^2+s^2}+r}{2}}
 +i\sgn(s)\sqrt{\frac{\sqrt{r^2+s^2}-r}{2}}.
 \label{eq:complex_root}
\end{equation}
A real propagating mode requires $s=0$ and $r>0$. With
\begin{equation}
 K=\frac{8}{[(\bar{\sigma}')^2+(\bar{\sigma}'')^2]^2},
\end{equation}
the two exact solutions of $s=0$ are
\begin{equation}
 \begin{aligned}
 \varepsilon_{d,\pm}''={}
 \frac{1-K\varepsilon_d'
 [(\bar{\sigma}')^2-(\bar{\sigma}'')^2]}
 {2K\bar{\sigma}'\bar{\sigma}''}
 {}\pm\frac{1}{2K\bar{\sigma}'\bar{\sigma}''}
 \Biggl(
 \bigl\{1-K\varepsilon_d'
 [(\bar{\sigma}')^2-(\bar{\sigma}'')^2]\bigr\}^2
 +4K^2(\varepsilon_d')^2
 (\bar{\sigma}')^2(\bar{\sigma}'')^2
 \Biggr)^{1/2}.
 \end{aligned}
 \label{eq:slg_exact_gain}
\end{equation}
For the parameters considered here, $\varepsilon_{d,-}''$ is negative and is the
gain-compensation root. The other root is lossy. In the strong-confinement limit
$|q|\gg k_0$, Eq.~\eqref{eq:slg_exact_gain} reduces to
\begin{equation}
 \varepsilon_{d,-}''
 \simeq-\varepsilon_d'\frac{\bar{\sigma}'}{\bar{\sigma}''}
 =-\varepsilon_d'\frac{\sigma'}{\sigma''}.
 \label{eq:slg_qs_gain}
\end{equation}
Equation~\eqref{eq:slg_qs_gain} is an electrostatic approximation, whereas all
full-EM curves below use Eq.~\eqref{eq:slg_exact_gain} or the unsquared boundary
condition \eqref{eq:slg_general}.

The separate boundary $r=0$ is
\begin{equation}
 \begin{aligned}
 \varepsilon_{d,c,\pm}''={}
 \frac{2\varepsilon_d'\bar{\sigma}'\bar{\sigma}''}
 {(\bar{\sigma}')^2-(\bar{\sigma}'')^2}
 {}\pm
 \frac{(\bar{\sigma}')^2+(\bar{\sigma}'')^2}
 {(\bar{\sigma}')^2-(\bar{\sigma}'')^2} \times
 \left\{(\varepsilon_d')^2-
 \frac{\varepsilon_d'}{4}
 [(\bar{\sigma}')^2-(\bar{\sigma}'')^2]\right\}^{1/2}.
 \end{aligned}
 \label{eq:r_boundary}
\end{equation}
These values identify a sign change of the real part of $R$; they are not, in
general, loss-compensation solutions because they need not satisfy $s=0$.

\begin{figure}[!htbp]
 \centering
 \includegraphics[width=0.92\linewidth]{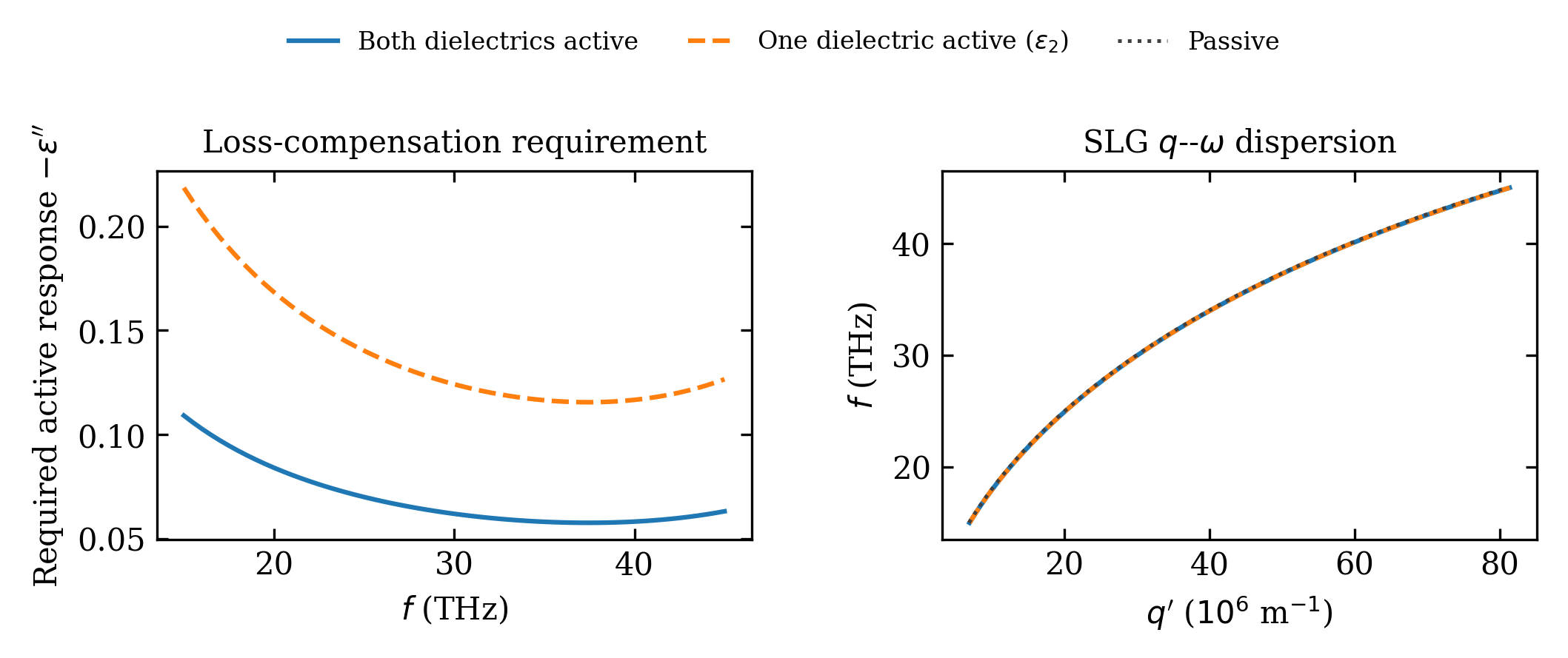}
 \caption{Full-EM SLG compensation for $E_F=0.20~\mathrm{eV}$,
$\tau=100~\mathrm{fs}$, $T=300~\mathrm{K}$, and
$\varepsilon_1'=\varepsilon_2'=1$. Left: required active response for symmetric
gain and for gain in only $\varepsilon_2$. Right: the corresponding real
$q$--$\omega$ dispersion; the compensated curves nearly coincide with the
passive real branch.}
 \label{fig:slg_gain_location}
\end{figure}

The symmetric SLG geometry is not the only possible gain distribution. For one
passive dielectric and one active dielectric, Eq.~\eqref{eq:slg_general} is
solved for the two real unknowns $q'$ and $\varepsilon_2''$ under the constraint
$q''=0$. At 30~THz, the required gain magnitudes are $-\varepsilon''=0.06211$
when both dielectrics are active and $-\varepsilon_2''=0.12422$ when only one
dielectric is active. The nearly twofold increase reflects the reduction of
active modal overlap, while the real dispersion changes negligibly. The
frequency dependence of both compensation conditions is shown in
Fig.~\ref{fig:slg_gain_location}.

The same full-EM boundary condition applies when the two claddings have
different real permittivities. Appendix~\ref{app:unequal_dielectrics} gives the
corresponding exact root problem and its closed strong-confinement limit.

\FloatBarrier
\section{Double-layer graphene}

Two identical graphene sheets are placed at $z=\pm d/2$. The upper and lower
half-spaces have $\varepsilon_1=\varepsilon_3$, and the spacer has permittivity
$\varepsilon_2$. Coupling of the two sheets produces optical and acoustic
plasmon branches \cite{profumo2012double,principi2014plasmon}. In each region
the TM magnetic field satisfies $(\partial_z^2-\kappa_j^2)H_y=0$. For $z>d/2$ it
can be written as
\begin{equation}
 H_y^{(1)}=B\exp[-\kappa_1(z-d/2)].
\end{equation}
The spacer fields may be chosen according to their parity,
\begin{equation}
 H_{y,\mathrm{opt}}^{(2)}=A\sinh(\kappa_2z),\qquad
 H_{y,\mathrm{ac}}^{(2)}=A\cosh(\kappa_2z).
 \label{eq:dlg_fields}
\end{equation}
The odd magnetic field gives equal longitudinal currents on the two sheets and
hence the in-phase-charge optical branch; the even magnetic field gives the
out-of-phase-charge acoustic branch. Since
\begin{equation}
 E_x=\frac{1}{i\omega\varepsilon_0\varepsilon_j}
 \frac{\partial H_y}{\partial z},
\end{equation}
continuity of $E_x$ and the jump condition $\hat{\bm z}\times(\bm H_1-\bm
H_2)=\bm J_s=\sigma\bm E_t$ at $z=d/2$ give
\begin{align}
 F_{\mathrm{opt}}={}&\frac{\varepsilon_2}{\kappa_2}
 \tanh\!\left(\frac{\kappa_2d}{2}\right)
 +\frac{\varepsilon_1}{\kappa_1}
 +i\frac{\sigma}{\omega\varepsilon_0}=0,
 \label{eq:dlg_opt}\\
 F_{\mathrm{ac}}={}&\frac{\varepsilon_2}{\kappa_2}
 \coth\!\left(\frac{\kappa_2d}{2}\right)
 +\frac{\varepsilon_1}{\kappa_1}
 +i\frac{\sigma}{\omega\varepsilon_0}=0.
 \label{eq:dlg_ac}
\end{align}
By symmetry, the boundary conditions at the lower sheet give the same dispersion
relations. The different hyperbolic factors arise from the optical and acoustic
field parities and describe the coupling between the sheets through the spacer.

For each mode $m\in\{\mathrm{opt},\mathrm{ac}\}$, exact full-EM compensation is
obtained by taking $q$ real and solving
\begin{equation}
 \operatorname{Re}F_m(q',\varepsilon_g'')=0,\qquad
 \operatorname{Im}F_m(q',\varepsilon_g'')=0
 \label{eq:dlg_compensation}
\end{equation}
for $q'$ and the imaginary part $\varepsilon_g''<0$ assigned to the active
dielectric region or regions. Because $\varepsilon_g''$ occurs inside $\kappa_j$
and the hyperbolic functions, Eq.~\eqref{eq:dlg_compensation} is transcendental
and has no general algebraic solution. All DLG results below are full-EM roots
of Eqs.~\eqref{eq:dlg_opt} and \eqref{eq:dlg_ac}; the strong-confinement limit
is given separately in Appendix~\ref{app:unequal_dielectrics}.

The appendix also gives the exact determinant for unequal upper and lower
claddings, for which optical and acoustic parity is no longer exact.

\begin{figure}[!htbp]
 \centering
 \includegraphics[width=0.94\linewidth]{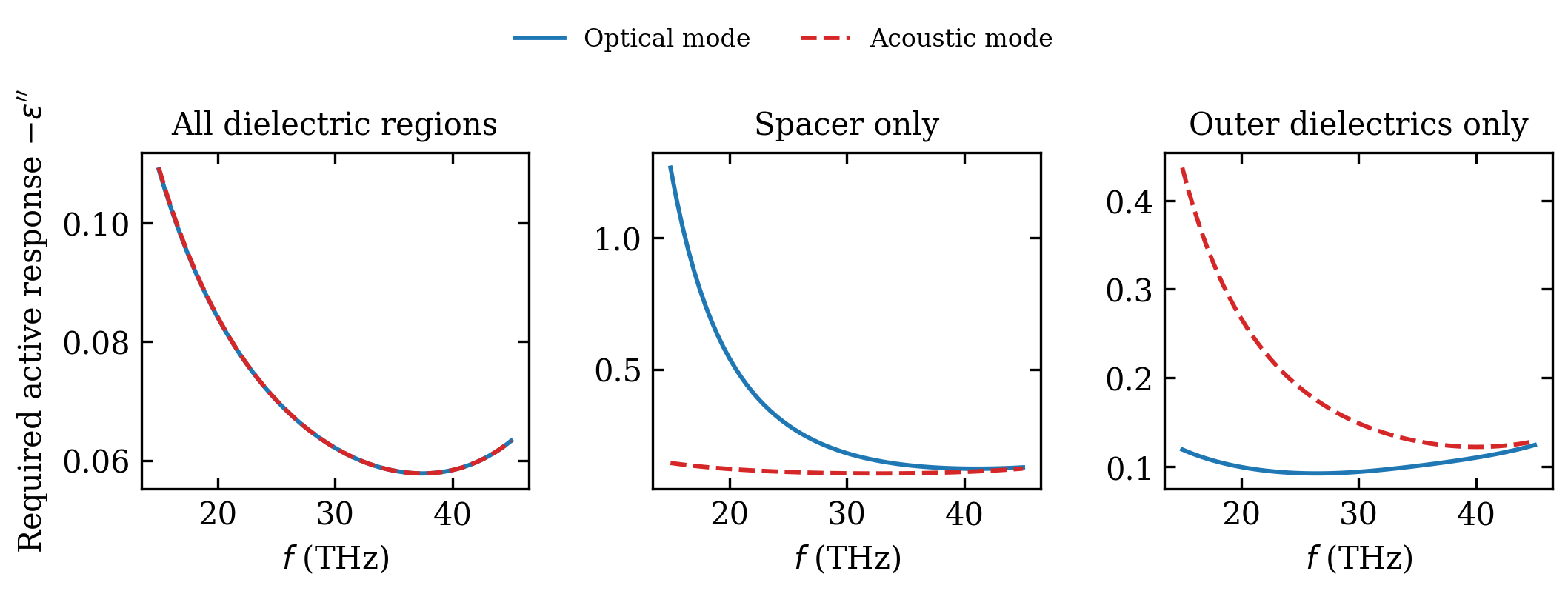}
 \caption{Branch-resolved full-EM DLG compensation for $d=50~\mathrm{nm}$,
$E_F=0.20~\mathrm{eV}$, $\tau=100~\mathrm{fs}$, $T=300~\mathrm{K}$, and
$\varepsilon_1'=\varepsilon_2'=\varepsilon_3'=1$. The optical and acoustic
thresholds nearly coincide when all dielectrics are active, but separate for
spacer-only and outer-only gain because the modal overlaps are different.}
 \label{fig:dlg_gain_location}
\end{figure}

We use $d=50~\mathrm{nm}$ and $\varepsilon_1'=\varepsilon_2'=\varepsilon_3'=1$.
Three gain distributions are compared: all dielectric regions active, only the
spacer active, and only the two outer dielectrics active. At 30~THz, uniform
gain requires $-\varepsilon_g''\simeq0.06211$ for both modes. Spacer-only gain
instead requires $0.1828$ for the optical mode and $0.1066$ for the acoustic
mode, whereas outer-only gain requires $0.09407$ and $0.1488$, respectively. The
acoustic mode therefore has the larger spacer overlap and the optical mode the
larger outer-medium overlap. Figure~\ref{fig:dlg_gain_location} shows this
branch-selective response over the full frequency range.

\FloatBarrier
\section{Local-response range}

Large values of $q/k_0$ quantify electromagnetic confinement but do not by
themselves determine the validity of a local graphene conductivity. More
relevant measures are $|q|/k_F$ and $v_F|q|/\omega$, where $k_F=E_F/(\hbar v_F)$
\cite{hwang2007dielectric}. Across 15--45~THz, the compensated SLG and DLG
branches satisfy $0.013\leq |q|/k_F\leq0.272$ and $0.040\leq
v_F|q|/\omega\leq0.292$. These values support the local Kubo treatment over the
chosen range, while their approach to $0.3$ at the upper end indicates that
nonlocality may produce quantitative corrections there.

\FloatBarrier
\section{Full-wave simulations}

\begin{figure}[!htbp]
 \centering
 \includegraphics{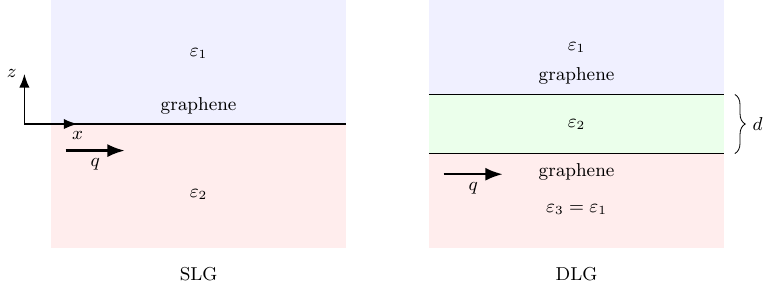}
 \caption{SLG and DLG dielectric assignments. The DLG sheets are separated by
the physical spacer thickness $d=50~\mathrm{nm}$; $\varepsilon_1$ and
$\varepsilon_3$ are the symmetric outer media and $\varepsilon_2$ is the spacer.
In COMSOL each graphene sheet is a $d_{\mathrm{eff}}=0.10~\mathrm{nm}$
conductive domain with $\sigma_{3\mathrm{D}}=\sigma/d_{\mathrm{eff}}$.}
 \label{fig:geometry}
\end{figure}

To compare the analytical results with a full-wave numerical treatment, we
perform frequency-domain FEM eigenmode calculations in COMSOL. The dielectric
assignments are summarized in Fig.~\ref{fig:geometry}. Graphene, treated
analytically as a zero-thickness sheet, is represented numerically by a thin
domain of effective thickness $d_{\mathrm{eff}}=0.10~\mathrm{nm}$ and bulk
conductivity $\sigma_{3\mathrm{D}}=\sigma/d_{\mathrm{eff}}$. A convergence check
at 30~THz shows that reducing $d_{\mathrm{eff}}$ from $0.345$ to
$0.10~\mathrm{nm}$ lowers the residual $|q''|$ by approximately a factor of
$3.5$ for both DLG branches, consistent with near-linear convergence toward the
zero-thickness limit. The residual attenuation is therefore consistent with the
finite-thickness numerical representation; the analytically compensated mode has
$q''=0$.

\FloatBarrier

COMSOL uses the opposite harmonic convention, $\exp(+i\omega t)$. We therefore
use the complex-conjugated conductivity and dielectric response relative to
Eqs.~\eqref{eq:conventions}--\eqref{eq:dlg_compensation}; a positive imaginary
permittivity in COMSOL corresponds to the negative $\varepsilon''$ used here.
The branches are identified by frequency continuation and their field parity.

\begin{figure}[!htb]
 \centering
 \includegraphics[width=0.88\linewidth]{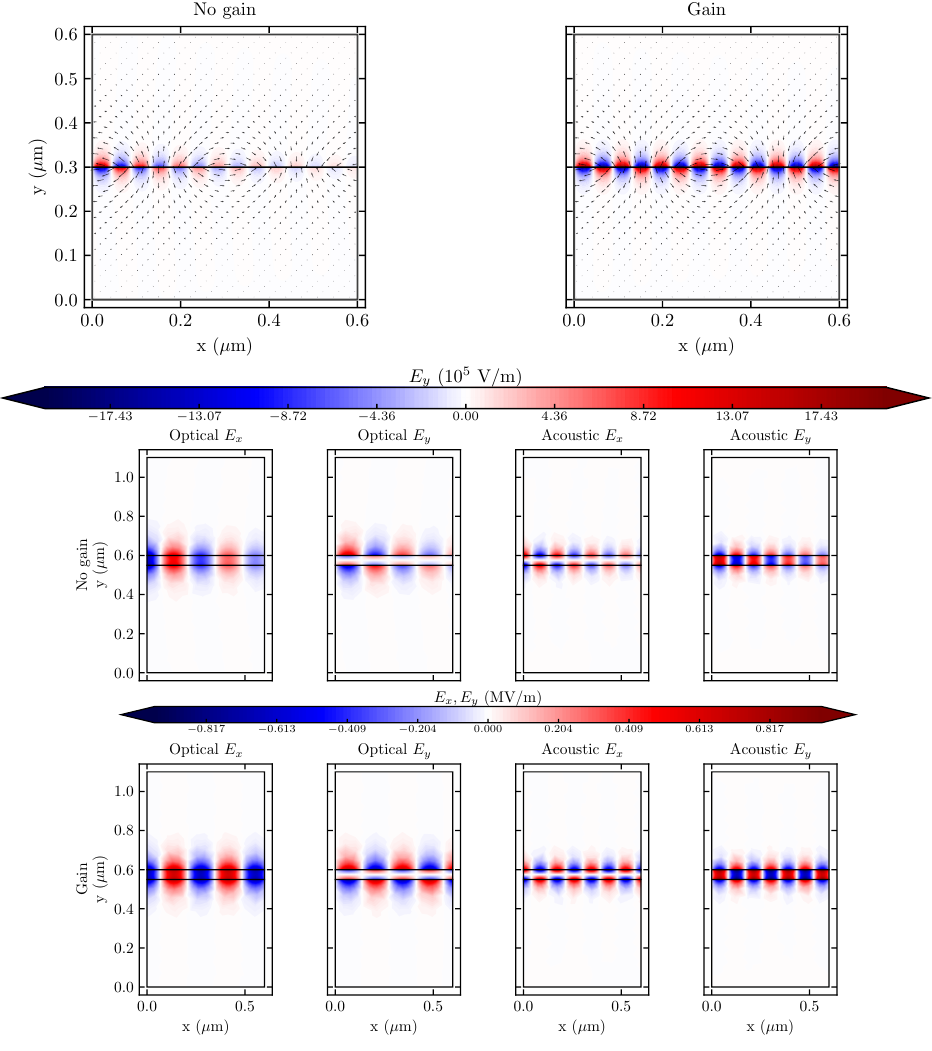}
 \caption{COMSOL field maps for passive and uniformly compensated graphene
structures. The upper pair shows the SLG field at 42~THz, including the in-plane
field direction; the lower array shows the optical and acoustic DLG $E_x$ and
$E_y$ components at 30~THz. The fields remain localized at the graphene sheets,
while the DLG components identify the two modal parities. The simulation
coordinate $y$ corresponds to $z$ in the analytical derivation. Here
$E_F=0.20~\mathrm{eV}$, $\tau=100~\mathrm{fs}$, $T=300~\mathrm{K}$,
$d=50~\mathrm{nm}$ for DLG, and all real dielectric permittivities are unity.}
 \label{fig:field_maps}
\end{figure}

\begin{figure}[!htb]
 \centering
 \includegraphics[width=0.78\linewidth]{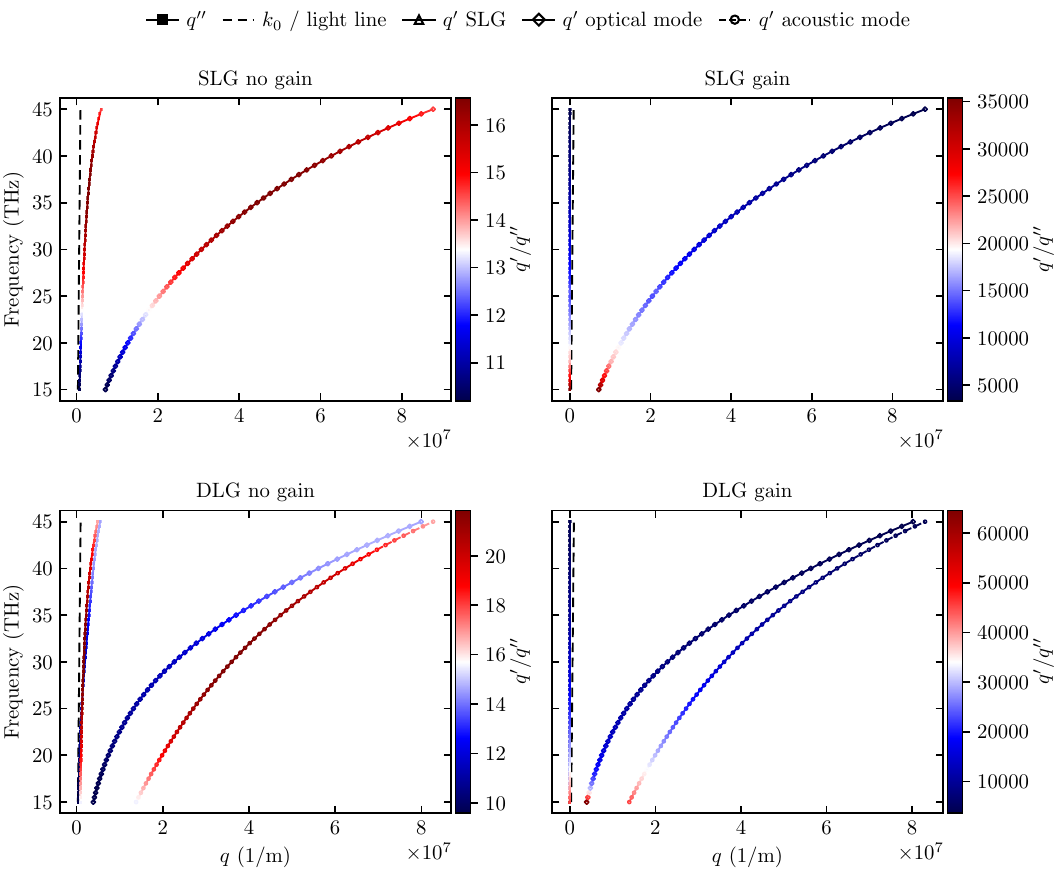}
 \caption{COMSOL $q$--$\omega$ dispersion for SLG and DLG over 15--45~THz. The
upper panels contain the SLG branch and the lower panels the optical and
acoustic DLG branches; color denotes $q'/|q''|$, and the light line is included
for reference. Here $E_F=0.20~\mathrm{eV}$, $\tau=100~\mathrm{fs}$,
$T=300~\mathrm{K}$, $d=50~\mathrm{nm}$ for DLG, and the real dielectric
permittivities are unity.}
 \label{fig:dispersion_comsol}
\end{figure}

Figure~\ref{fig:field_maps} shows the confined field profiles associated with
the computed graphene SPP eigenvalues. The optical and acoustic DLG maps exhibit
the parities introduced in Eq.~\eqref{eq:dlg_fields}. The compensated modes
retain their branch character: gain modifies the attenuation without creating an
additional eigenmode. The corresponding SLG and DLG $q$--$\omega$ branches are
shown in Fig.~\ref{fig:dispersion_comsol}.

\FloatBarrier

\begin{figure}[!htb]
 \centering
 \includegraphics[width=0.78\linewidth]{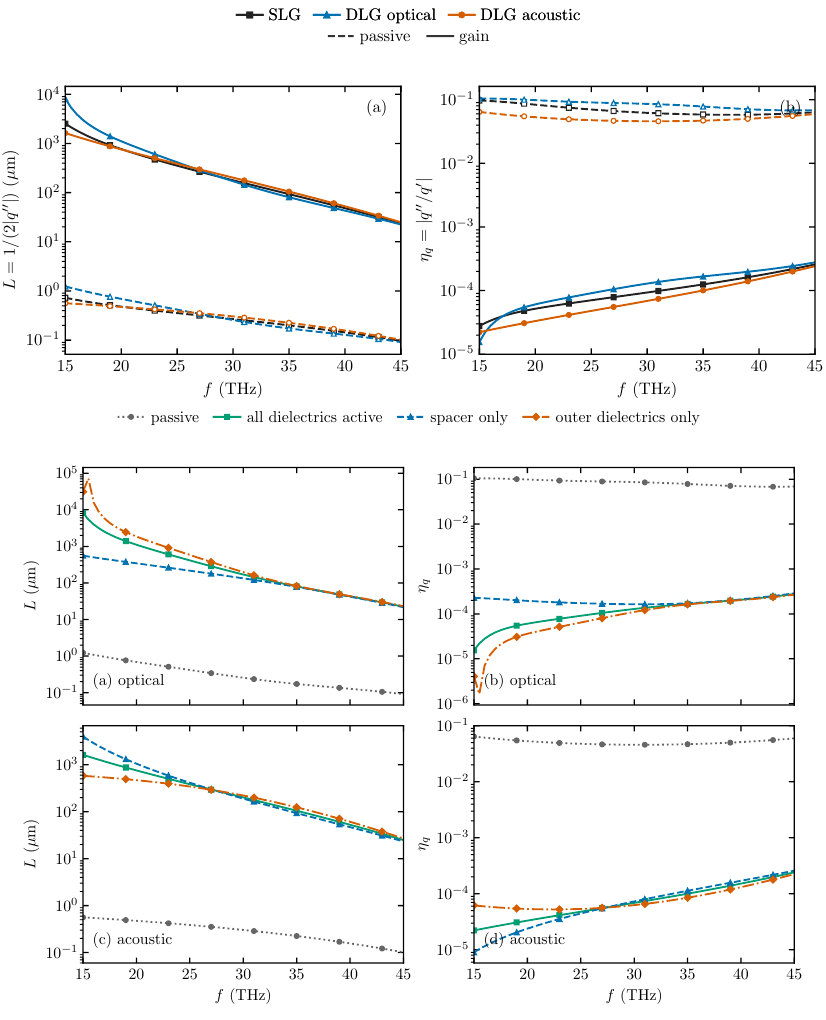}
 \caption{FEM propagation diagnostics over 15--45~THz. The upper block compares
$L=1/(2|q''|)$ and $\eta_q=|q''/q'|$ for passive and uniformly compensated SLG
and DLG. The lower block resolves the DLG optical and acoustic modes for
uniform, spacer-only, and outer-dielectric gain, with the appropriate threshold
used for each branch and distribution. Narrow maxima in $L$ mark residual
numerical $q''$ values closest to zero and are not additional resonances. Here
$d=50~\mathrm{nm}$, $E_F=0.20~\mathrm{eV}$, $\tau=100~\mathrm{fs}$, and
$T=300~\mathrm{K}$.}
 \label{fig:propagation_results}
\end{figure}

To quantify compensation independently of the sign convention, we use
\begin{equation}
 \eta_q=\left|\frac{q''}{q'}\right|.
 \label{eq:eta}
\end{equation}
At 30~THz, the FEM intensity propagation length $L=1/(2|q''|)$ increases from
$0.27$ to $178~\mu\mathrm{m}$ for SLG, from $0.26$ to $170~\mu\mathrm{m}$ for
the optical DLG mode, and from $0.30$ to $202~\mu\mathrm{m}$ for the acoustic
mode. These changes correspond to an approximately three-order-of-magnitude
reduction in $\eta_q$. The finite compensated lengths are lower bounds inferred
from the residual FEM attenuation; the analytical full-EM solutions have $q''=0$
and hence $L\rightarrow\infty$ within the local linear model. The upper block of
Fig.~\ref{fig:propagation_results} summarizes the propagation-length enhancement
and attenuation reduction.

Uniform compensation over 15--45~THz requires $-\varepsilon''=0.058$--$0.109$
for all three branches, corresponding to the material intensity-gain coefficient
\begin{equation}
 g_{\mathrm{mat}}=-2k_0\operatorname{Im}\sqrt{\varepsilon'+i\varepsilon''}
\end{equation}
of approximately $(3.4$--$6.0)\times10^2~\mathrm{cm^{-1}}$. At 30~THz,
$-\varepsilon''\simeq0.0621$ and
$g_{\mathrm{mat}}\simeq3.9\times10^2~\mathrm{cm^{-1}}$. These are material
requirements of the linear model; practical implementations must also satisfy
bandwidth, saturation, and integration constraints
\cite{yu2008gain,athanasopoulos2013enhanced,oulton2009plasmon}.

The placement-resolved FEM sweeps use the separate full-EM compensation
threshold calculated for each gain distribution and, for DLG, for each branch.
The analogous SLG calculations with gain in both dielectrics and in one
dielectric reach essentially the same residual attenuation when their respective
thresholds are applied, reproducing the single-branch trend already established
in Fig.~\ref{fig:slg_gain_location}; they are therefore not shown separately.
The lower block of Fig.~\ref{fig:propagation_results} gives the branch-resolved
DLG results. At 30~THz the compensated DLG calculations give
$6.3\times10^{-5}\leq\eta_q\leq1.6\times10^{-4}$ across all three gain
placements, compared with $8.6\times10^{-2}$ and $4.6\times10^{-2}$ for the
passive optical and acoustic modes. The small spread among the finite
compensated values reflects the effective-thickness representation, mesh, and
eigensolver tolerance; it is not interpreted as a physical preference for one
gain placement in the ideal $q''=0$ model.

\FloatBarrier
\section{Conclusions}

We have formulated gain compensation for SLG and DLG within one full-EM,
local-conductivity framework. For symmetric SLG, the exact effective-index
equation provides an algebraic compensation root; the frequently used expression
$\varepsilon''\simeq-\varepsilon'\sigma'/\sigma''$ is recovered only in the
strong-confinement limit. For DLG, the optical and acoustic equations follow
from the odd and even spacer-field parities and must, in general, be solved as
separate transcendental problems.

The DLG analysis reveals a mode-dependent response to gain placement. The
optical and acoustic thresholds are almost identical when every dielectric
region is active, but separate when gain is confined to the spacer or to the
outer dielectrics because the two modes have different spatial overlaps.
Full-wave calculations reproduce both branches and reduce $|q''/q'|$ by
approximately three orders of magnitude relative to the passive structures. The
residual FEM attenuation decreases toward the zero-thickness limit, whereas the
analytical full-EM roots satisfy $q''=0$ directly.

The locality diagnostics remain below $0.3$ over the frequency range studied,
but approach that value at the upper end. Nonlocal conductivity and gain
saturation are therefore natural extensions for quantitative device modeling
beyond the present local, linear theory.

\section*{Acknowledgments}
This research was co-funded by the Stavros Niarchos Foundation (SNF) and the
Hellenic Foundation for Research and Innovation (H.F.R.I.) under the 5th Call of
Science and Society Action---Always Strive for Excellence---Theodore Papazoglou
(Project Number: 011496).

\appendix
\section{Unequal dielectric environments}
\label{app:unequal_dielectrics}

\subsection{Single-layer graphene}

Equation~\eqref{eq:slg_general} remains valid without assuming
$\varepsilon_1=\varepsilon_2$. With $n=q/k_0$,
$\gamma_j=\sqrt{n^2-\varepsilon_j}$, and $\bar\sigma=Z_0\sigma$, the exact
dimensionless equation is
\begin{equation}
 \mathcal F_{\mathrm{SLG}}(n)=
 \frac{\varepsilon_1}{\gamma_1}
 +\frac{\varepsilon_2}{\gamma_2}
 +i\bar\sigma=0.
 \label{eq:app_slg_exact}
\end{equation}
For a specified distribution of dielectric gain, full-EM compensation is
obtained by taking $n$ real and solving $\operatorname{Re}\mathcal
F_{\mathrm{SLG}}=\operatorname{Im}\mathcal F_{\mathrm{SLG}}=0$ for $n$ and the
gain amplitude. Unlike the symmetric result in Eq.~\eqref{eq:slg_index},
Eq.~\eqref{eq:app_slg_exact} does not generally reduce to a useful algebraic
expression.

In the strong-confinement limit, $|n|\gg\sqrt{|\varepsilon_j|}$ and
$\gamma_j\simeq n$, so
\begin{equation}
 n\simeq i\frac{\varepsilon_1+\varepsilon_2}{\bar\sigma}.
 \label{eq:app_slg_qs}
\end{equation}
Writing $\varepsilon_j=\varepsilon_j'+i\varepsilon_j''$ and requiring $n$ to be
real gives
\begin{align}
 \varepsilon_1''+\varepsilon_2''
 &=-\left(\varepsilon_1'+\varepsilon_2'\right)
 \frac{\sigma'}{\sigma''},
 \label{eq:app_slg_gain_sum}\\
 q'&=\frac{\omega\varepsilon_0
 (\varepsilon_1'+\varepsilon_2')}{\sigma''}.
 \label{eq:app_slg_qreal}
\end{align}
More generally, if $\varepsilon_j''=w_j\varepsilon_g''$ with prescribed weights
$w_j$, then
\begin{equation}
 \varepsilon_g''=-\frac{\varepsilon_1'+\varepsilon_2'}{w_1+w_2}
 \frac{\sigma'}{\sigma''}.
 \label{eq:app_slg_weighted}
\end{equation}
Thus one active cladding corresponds to $(w_1,w_2)=(0,1)$, whereas equal gain in
both claddings corresponds to $(1,1)$.
Equations~\eqref{eq:app_slg_gain_sum}--\eqref{eq:app_slg_weighted} are
strong-confinement limits; retarded values follow from
Eq.~\eqref{eq:app_slg_exact}.

\subsection{Double-layer graphene with symmetric outer media}

Let $\varepsilon_{\mathrm{out}}=\varepsilon_1=\varepsilon_3$ and
$\varepsilon_{\mathrm{sp}}=\varepsilon_2$, without requiring their real parts to
be equal. In the strong-confinement limit, Eqs.~\eqref{eq:dlg_opt} and
\eqref{eq:dlg_ac} become
\begin{equation}
 q_m\simeq i\frac{\omega\varepsilon_0}{\sigma}
 \left[\varepsilon_{\mathrm{out}}
 +T_m(q_m d)\varepsilon_{\mathrm{sp}}\right],
 \label{eq:app_dlg_qs}
\end{equation}
where
\begin{equation}
 T_{\mathrm{opt}}=\tanh\!\left(\frac{q_m d}{2}\right),
 \qquad
 T_{\mathrm{ac}}=\coth\!\left(\frac{q_m d}{2}\right).
 \label{eq:app_dlg_T}
\end{equation}
For a compensated mode, $q_m=q_m'$ and $T_m$ is real. Separating
Eq.~\eqref{eq:app_dlg_qs} into real and imaginary parts gives the coupled
relations
\begin{align}
 q_m'&=\frac{\omega\varepsilon_0}{\sigma''}
 \left(\varepsilon_{\mathrm{out}}'
 +T_m\varepsilon_{\mathrm{sp}}'\right),
 \label{eq:app_dlg_qreal}\\
 \varepsilon_{\mathrm{out}}''+T_m\varepsilon_{\mathrm{sp}}''
 &=-\left(\varepsilon_{\mathrm{out}}'
 +T_m\varepsilon_{\mathrm{sp}}'\right)
 \frac{\sigma'}{\sigma''}.
 \label{eq:app_dlg_gain_general}
\end{align}
The first equation determines $q_m'$ implicitly because $T_m$ depends on it. The
second equation then gives the gain required for that branch. For the three gain
distributions considered in the main text, Eq.~\eqref{eq:app_dlg_gain_general}
yields
\begin{align}
 \varepsilon_{g,m}''\big|_{\mathrm{all}}
 &=-\frac{\varepsilon_{\mathrm{out}}'
 +T_m\varepsilon_{\mathrm{sp}}'}{1+T_m}
 \frac{\sigma'}{\sigma''},
 \label{eq:app_dlg_all}\\
 \varepsilon_{g,m}''\big|_{\mathrm{sp}}
 &=-\frac{\varepsilon_{\mathrm{out}}'
 +T_m\varepsilon_{\mathrm{sp}}'}{T_m}
 \frac{\sigma'}{\sigma''},
 \label{eq:app_dlg_sp}\\
 \varepsilon_{g,m}''\big|_{\mathrm{out}}
 &=-\left(\varepsilon_{\mathrm{out}}'
 +T_m\varepsilon_{\mathrm{sp}}'\right)
 \frac{\sigma'}{\sigma''}.
 \label{eq:app_dlg_out}
\end{align}
These formulas make the branch dependence explicit: the optical and acoustic
modes sample the spacer through different values of $T_m$. For
$\varepsilon_{\mathrm{out}}'=\varepsilon_{\mathrm{sp}}'$,
Eq.~\eqref{eq:app_dlg_all} reduces to Eq.~\eqref{eq:slg_qs_gain}, while the
spacer-only and outer-only expressions retain their mode dependence. The full-EM
values are obtained from Eqs.~\eqref{eq:dlg_opt}--\eqref{eq:dlg_compensation}
rather than from the strong-confinement approximation.

\subsection{Fully asymmetric double-layer environment}

When $\varepsilon_1\neq\varepsilon_3$, reflection symmetry is absent and the
modes do not possess exact optical or acoustic parity. The two-sheet full-EM
problem can nevertheless be written as a single determinant. For sheet
conductivities $\sigma_1$ and $\sigma_2$, define
\begin{align}
 A_1={}&\frac{\varepsilon_1}{\kappa_1}
 +\frac{\varepsilon_2}{\kappa_2}\coth(\kappa_2d)
 +i\frac{\sigma_1}{\omega\varepsilon_0},\\
 A_3={}&\frac{\varepsilon_3}{\kappa_3}
 +\frac{\varepsilon_2}{\kappa_2}\coth(\kappa_2d)
 +i\frac{\sigma_2}{\omega\varepsilon_0},\\
 C={}&\frac{\varepsilon_2}{\kappa_2}\operatorname{csch}(\kappa_2d).
\end{align}
The dispersion relation is
\begin{equation}
 \mathcal D(q)=A_1A_3-C^2=0.
 \label{eq:app_dlg_determinant}
\end{equation}
For identical sheets and symmetric outer media,
Eq.~\eqref{eq:app_dlg_determinant} factorizes by using $\coth
x-\operatorname{csch}x=\tanh(x/2)$ and $\coth
x+\operatorname{csch}x=\coth(x/2)$, recovering Eqs.~\eqref{eq:dlg_opt} and
\eqref{eq:dlg_ac}. For unequal outer media, compensated solutions are found by
setting $q=q'$ and solving
\begin{equation}
 \operatorname{Re}\mathcal D(q',\varepsilon_g'')=0,
 \qquad
 \operatorname{Im}\mathcal D(q',\varepsilon_g'')=0,
 \label{eq:app_dlg_asym_comp}
\end{equation}
after prescribing how the single gain amplitude $\varepsilon_g''$ is distributed
among the dielectric regions. The two solutions may be followed continuously
from the symmetric limit as optical-like and acoustic-like branches, but those
labels no longer denote exact parity.


\begin{thebibliography}{35}%
\makeatletter
\providecommand \@ifxundefined [1]{%
 \@ifx{#1\undefined}
}%
\providecommand \@ifnum [1]{%
 \ifnum #1\expandafter \@firstoftwo
 \else \expandafter \@secondoftwo
 \fi
}%
\providecommand \@ifx [1]{%
 \ifx #1\expandafter \@firstoftwo
 \else \expandafter \@secondoftwo
 \fi
}%
\providecommand \natexlab [1]{#1}%
\providecommand \enquote  [1]{``#1''}%
\providecommand \bibnamefont  [1]{#1}%
\providecommand \bibfnamefont [1]{#1}%
\providecommand \citenamefont [1]{#1}%
\providecommand \href@noop [0]{\@secondoftwo}%
\providecommand \href [0]{\begingroup \@sanitize@url \@href}%
\providecommand \@href[1]{\@@startlink{#1}\@@href}%
\providecommand \@@href[1]{\endgroup#1\@@endlink}%
\providecommand \@sanitize@url [0]{\catcode `\\12\catcode `\$12\catcode
  `\&12\catcode `\#12\catcode `\^12\catcode `\_12\catcode `\%12\relax}%
\providecommand \@@startlink[1]{}%
\providecommand \@@endlink[0]{}%
\providecommand \url  [0]{\begingroup\@sanitize@url \@url }%
\providecommand \@url [1]{\endgroup\@href {#1}{\urlprefix }}%
\providecommand \urlprefix  [0]{URL }%
\providecommand \Eprint [0]{\href }%
\providecommand \doibase [0]{https://doi.org/}%
\providecommand \selectlanguage [0]{\@gobble}%
\providecommand \bibinfo  [0]{\@secondoftwo}%
\providecommand \bibfield  [0]{\@secondoftwo}%
\providecommand \translation [1]{[#1]}%
\providecommand \BibitemOpen [0]{}%
\providecommand \bibitemStop [0]{}%
\providecommand \bibitemNoStop [0]{.\EOS\space}%
\providecommand \EOS [0]{\spacefactor3000\relax}%
\providecommand \BibitemShut  [1]{\csname bibitem#1\endcsname}%
\let\auto@bib@innerbib\@empty
\bibitem [{\citenamefont {Ritchie}(1957)}]{ritchie1957plasma}%
  \BibitemOpen
  \bibfield  {author} {\bibinfo {author} {\bibfnamefont {R.~H.}\ \bibnamefont
  {Ritchie}},\ }\bibfield  {title} {\bibinfo {title} {Plasma losses by fast
  electrons in thin films},\ }\href@noop {} {\bibfield  {journal} {\bibinfo
  {journal} {Physical Review}\ }\textbf {\bibinfo {volume} {106}},\ \bibinfo
  {pages} {874} (\bibinfo {year} {1957})}\BibitemShut {NoStop}%
\bibitem [{\citenamefont {Economou}(1969)}]{economou1969surface}%
  \BibitemOpen
  \bibfield  {author} {\bibinfo {author} {\bibfnamefont {E.~N.}\ \bibnamefont
  {Economou}},\ }\bibfield  {title} {\bibinfo {title} {Surface plasmons in thin
  films},\ }\href {https://doi.org/10.1103/PhysRev.182.539} {\bibfield
  {journal} {\bibinfo  {journal} {Phys. Rev.}\ }\textbf {\bibinfo {volume}
  {182}},\ \bibinfo {pages} {539} (\bibinfo {year} {1969})}\BibitemShut
  {NoStop}%
\bibitem [{\citenamefont {Pitarke}\ \emph {et~al.}(2006)\citenamefont
  {Pitarke}, \citenamefont {Silkin}, \citenamefont {Chulkov},\ and\
  \citenamefont {Echenique}}]{pitarke2006theory}%
  \BibitemOpen
  \bibfield  {author} {\bibinfo {author} {\bibfnamefont {J.}~\bibnamefont
  {Pitarke}}, \bibinfo {author} {\bibfnamefont {V.}~\bibnamefont {Silkin}},
  \bibinfo {author} {\bibfnamefont {E.}~\bibnamefont {Chulkov}},\ and\ \bibinfo
  {author} {\bibfnamefont {P.}~\bibnamefont {Echenique}},\ }\bibfield  {title}
  {\bibinfo {title} {Theory of surface plasmons and surface-plasmon
  polaritons},\ }\href@noop {} {\bibfield  {journal} {\bibinfo  {journal}
  {Reports on Progress in Physics}\ }\textbf {\bibinfo {volume} {70}},\
  \bibinfo {pages} {1} (\bibinfo {year} {2006})}\BibitemShut {NoStop}%
\bibitem [{\citenamefont {Barnes}\ \emph {et~al.}(2003)\citenamefont {Barnes},
  \citenamefont {Dereux},\ and\ \citenamefont {Ebbesen}}]{barnes2003surface}%
  \BibitemOpen
  \bibfield  {author} {\bibinfo {author} {\bibfnamefont {W.~L.}\ \bibnamefont
  {Barnes}}, \bibinfo {author} {\bibfnamefont {A.}~\bibnamefont {Dereux}},\
  and\ \bibinfo {author} {\bibfnamefont {T.~W.}\ \bibnamefont {Ebbesen}},\
  }\bibfield  {title} {\bibinfo {title} {Surface plasmon subwavelength
  optics},\ }\href@noop {} {\bibfield  {journal} {\bibinfo  {journal} {Nature}\
  }\textbf {\bibinfo {volume} {424}},\ \bibinfo {pages} {824} (\bibinfo {year}
  {2003})}\BibitemShut {NoStop}%
\bibitem [{\citenamefont {Ozbay}(2006)}]{ozbay2006plasmonics}%
  \BibitemOpen
  \bibfield  {author} {\bibinfo {author} {\bibfnamefont {E.}~\bibnamefont
  {Ozbay}},\ }\bibfield  {title} {\bibinfo {title} {Plasmonics: merging
  photonics and electronics at nanoscale dimensions},\ }\href@noop {}
  {\bibfield  {journal} {\bibinfo  {journal} {Science}\ }\textbf {\bibinfo
  {volume} {311}},\ \bibinfo {pages} {189} (\bibinfo {year}
  {2006})}\BibitemShut {NoStop}%
\bibitem [{\citenamefont {Schuller}\ \emph {et~al.}(2010)\citenamefont
  {Schuller}, \citenamefont {Barnard}, \citenamefont {Cai}, \citenamefont
  {Jun}, \citenamefont {White},\ and\ \citenamefont
  {Brongersma}}]{schuller2010plasmonics}%
  \BibitemOpen
  \bibfield  {author} {\bibinfo {author} {\bibfnamefont {J.~A.}\ \bibnamefont
  {Schuller}}, \bibinfo {author} {\bibfnamefont {E.~S.}\ \bibnamefont
  {Barnard}}, \bibinfo {author} {\bibfnamefont {W.}~\bibnamefont {Cai}},
  \bibinfo {author} {\bibfnamefont {Y.~C.}\ \bibnamefont {Jun}}, \bibinfo
  {author} {\bibfnamefont {J.~S.}\ \bibnamefont {White}},\ and\ \bibinfo
  {author} {\bibfnamefont {M.~L.}\ \bibnamefont {Brongersma}},\ }\bibfield
  {title} {\bibinfo {title} {Plasmonics for extreme light concentration and
  manipulation},\ }\href@noop {} {\bibfield  {journal} {\bibinfo  {journal}
  {Nature Materials}\ }\textbf {\bibinfo {volume} {9}},\ \bibinfo {pages} {193}
  (\bibinfo {year} {2010})}\BibitemShut {NoStop}%
\bibitem [{\citenamefont {Gramotnev}\ and\ \citenamefont
  {Bozhevolnyi}(2010)}]{gramotnev2010plasmonics}%
  \BibitemOpen
  \bibfield  {author} {\bibinfo {author} {\bibfnamefont {D.~K.}\ \bibnamefont
  {Gramotnev}}\ and\ \bibinfo {author} {\bibfnamefont {S.~I.}\ \bibnamefont
  {Bozhevolnyi}},\ }\bibfield  {title} {\bibinfo {title} {Plasmonics beyond the
  diffraction limit},\ }\href@noop {} {\bibfield  {journal} {\bibinfo
  {journal} {Nature Photonics}\ }\textbf {\bibinfo {volume} {4}},\ \bibinfo
  {pages} {83} (\bibinfo {year} {2010})}\BibitemShut {NoStop}%
\bibitem [{\citenamefont {Maier}(2007)}]{maier2007plasmonics}%
  \BibitemOpen
  \bibfield  {author} {\bibinfo {author} {\bibfnamefont {S.~A.}\ \bibnamefont
  {Maier}},\ }\href@noop {} {\emph {\bibinfo {title} {Plasmonics: fundamentals
  and applications}}}\ (\bibinfo  {publisher} {Springer},\ \bibinfo {year}
  {2007})\BibitemShut {NoStop}%
\bibitem [{\citenamefont {Han}\ and\ \citenamefont
  {Bozhevolnyi}(2012)}]{Han_2013}%
  \BibitemOpen
  \bibfield  {author} {\bibinfo {author} {\bibfnamefont {Z.}~\bibnamefont
  {Han}}\ and\ \bibinfo {author} {\bibfnamefont {S.~I.}\ \bibnamefont
  {Bozhevolnyi}},\ }\bibfield  {title} {\bibinfo {title} {Radiation guiding
  with surface plasmon polaritons},\ }\href
  {https://doi.org/10.1088/0034-4885/76/1/016402} {\bibfield  {journal}
  {\bibinfo  {journal} {Reports on Progress in Physics}\ }\textbf {\bibinfo
  {volume} {76}},\ \bibinfo {pages} {016402} (\bibinfo {year}
  {2012})}\BibitemShut {NoStop}%
\bibitem [{\citenamefont {Smajic}\ \emph {et~al.}(2009)\citenamefont {Smajic},
  \citenamefont {Hafner}, \citenamefont {Raguin}, \citenamefont
  {Tavzarashvili},\ and\ \citenamefont {Mishrikey}}]{smajic2009comparison}%
  \BibitemOpen
  \bibfield  {author} {\bibinfo {author} {\bibfnamefont {J.}~\bibnamefont
  {Smajic}}, \bibinfo {author} {\bibfnamefont {C.}~\bibnamefont {Hafner}},
  \bibinfo {author} {\bibfnamefont {L.}~\bibnamefont {Raguin}}, \bibinfo
  {author} {\bibfnamefont {K.}~\bibnamefont {Tavzarashvili}},\ and\ \bibinfo
  {author} {\bibfnamefont {M.}~\bibnamefont {Mishrikey}},\ }\bibfield  {title}
  {\bibinfo {title} {Comparison of numerical methods for the analysis of
  plasmonic structures},\ }\href@noop {} {\bibfield  {journal} {\bibinfo
  {journal} {Journal of Computational and Theoretical Nanoscience}\ }\textbf
  {\bibinfo {volume} {6}},\ \bibinfo {pages} {763} (\bibinfo {year}
  {2009})}\BibitemShut {NoStop}%
\bibitem [{\citenamefont {Jin}(2014)}]{jin2014finite}%
  \BibitemOpen
  \bibfield  {author} {\bibinfo {author} {\bibfnamefont {J.-M.}\ \bibnamefont
  {Jin}},\ }\href@noop {} {\emph {\bibinfo {title} {The Finite Element Method
  in Electromagnetics}}},\ \bibinfo {edition} {3rd}\ ed.\ (\bibinfo
  {publisher} {Wiley},\ \bibinfo {year} {2014})\BibitemShut {NoStop}%
\bibitem [{\citenamefont {Castro~Neto}\ \emph {et~al.}(2009)\citenamefont
  {Castro~Neto}, \citenamefont {Guinea}, \citenamefont {Peres}, \citenamefont
  {Novoselov},\ and\ \citenamefont {Geim}}]{castro2009electronic}%
  \BibitemOpen
  \bibfield  {author} {\bibinfo {author} {\bibfnamefont {A.~H.}\ \bibnamefont
  {Castro~Neto}}, \bibinfo {author} {\bibfnamefont {F.}~\bibnamefont {Guinea}},
  \bibinfo {author} {\bibfnamefont {N.~M.~R.}\ \bibnamefont {Peres}}, \bibinfo
  {author} {\bibfnamefont {K.~S.}\ \bibnamefont {Novoselov}},\ and\ \bibinfo
  {author} {\bibfnamefont {A.~K.}\ \bibnamefont {Geim}},\ }\bibfield  {title}
  {\bibinfo {title} {The electronic properties of graphene},\ }\href@noop {}
  {\bibfield  {journal} {\bibinfo  {journal} {Reviews of Modern Physics}\
  }\textbf {\bibinfo {volume} {81}},\ \bibinfo {pages} {109} (\bibinfo {year}
  {2009})}\BibitemShut {NoStop}%
\bibitem [{\citenamefont {Peres}(2010)}]{peres2010colloquium}%
  \BibitemOpen
  \bibfield  {author} {\bibinfo {author} {\bibfnamefont {N.~M.~R.}\
  \bibnamefont {Peres}},\ }\bibfield  {title} {\bibinfo {title} {Colloquium:
  The transport properties of graphene: An introduction},\ }\href@noop {}
  {\bibfield  {journal} {\bibinfo  {journal} {Reviews of Modern Physics}\
  }\textbf {\bibinfo {volume} {82}},\ \bibinfo {pages} {2673} (\bibinfo {year}
  {2010})}\BibitemShut {NoStop}%
\bibitem [{\citenamefont {Jablan}\ \emph {et~al.}(2009)\citenamefont {Jablan},
  \citenamefont {Buljan},\ and\ \citenamefont
  {Solja{\v{c}}i{\'c}}}]{jablan2009plasmonics}%
  \BibitemOpen
  \bibfield  {author} {\bibinfo {author} {\bibfnamefont {M.}~\bibnamefont
  {Jablan}}, \bibinfo {author} {\bibfnamefont {H.}~\bibnamefont {Buljan}},\
  and\ \bibinfo {author} {\bibfnamefont {M.}~\bibnamefont
  {Solja{\v{c}}i{\'c}}},\ }\bibfield  {title} {\bibinfo {title} {Plasmonics in
  graphene at infrared frequencies},\ }\href@noop {} {\bibfield  {journal}
  {\bibinfo  {journal} {Physical Review B}\ }\textbf {\bibinfo {volume} {80}},\
  \bibinfo {pages} {245435} (\bibinfo {year} {2009})}\BibitemShut {NoStop}%
\bibitem [{\citenamefont {Koppens}\ \emph {et~al.}(2011)\citenamefont
  {Koppens}, \citenamefont {Chang},\ and\ \citenamefont {Garc{\'\i}a~de
  Abajo}}]{koppens2011graphene}%
  \BibitemOpen
  \bibfield  {author} {\bibinfo {author} {\bibfnamefont {F.~H.}\ \bibnamefont
  {Koppens}}, \bibinfo {author} {\bibfnamefont {D.~E.}\ \bibnamefont {Chang}},\
  and\ \bibinfo {author} {\bibfnamefont {F.~J.}\ \bibnamefont {Garc{\'\i}a~de
  Abajo}},\ }\bibfield  {title} {\bibinfo {title} {Graphene plasmonics: a
  platform for strong light--matter interactions},\ }\href@noop {} {\bibfield
  {journal} {\bibinfo  {journal} {Nano Letters}\ }\textbf {\bibinfo {volume}
  {11}},\ \bibinfo {pages} {3370} (\bibinfo {year} {2011})}\BibitemShut
  {NoStop}%
\bibitem [{\citenamefont {Jablan}\ \emph {et~al.}(2013)\citenamefont {Jablan},
  \citenamefont {Solja{\v{c}}i{\'c}},\ and\ \citenamefont
  {Buljan}}]{jablan2013plasmons}%
  \BibitemOpen
  \bibfield  {author} {\bibinfo {author} {\bibfnamefont {M.}~\bibnamefont
  {Jablan}}, \bibinfo {author} {\bibfnamefont {M.}~\bibnamefont
  {Solja{\v{c}}i{\'c}}},\ and\ \bibinfo {author} {\bibfnamefont
  {H.}~\bibnamefont {Buljan}},\ }\bibfield  {title} {\bibinfo {title} {Plasmons
  in graphene: fundamental properties and potential applications},\ }\href@noop
  {} {\bibfield  {journal} {\bibinfo  {journal} {Proceedings of the IEEE}\
  }\textbf {\bibinfo {volume} {101}},\ \bibinfo {pages} {1689} (\bibinfo {year}
  {2013})}\BibitemShut {NoStop}%
\bibitem [{\citenamefont {Gon{\c{c}}alves}\ and\ \citenamefont
  {Peres}(2016)}]{gonccalves2016introduction}%
  \BibitemOpen
  \bibfield  {author} {\bibinfo {author} {\bibfnamefont {P.~A.~D.}\
  \bibnamefont {Gon{\c{c}}alves}}\ and\ \bibinfo {author} {\bibfnamefont
  {N.~M.}\ \bibnamefont {Peres}},\ }\href@noop {} {\emph {\bibinfo {title} {An
  introduction to graphene plasmonics}}}\ (\bibinfo  {publisher} {World
  Scientific},\ \bibinfo {year} {2016})\BibitemShut {NoStop}%
\bibitem [{\citenamefont {Fei}\ \emph {et~al.}(2012)\citenamefont {Fei},
  \citenamefont {Rodin}, \citenamefont {Andreev}, \citenamefont {Bao},
  \citenamefont {McLeod}, \citenamefont {Wagner}, \citenamefont {Zhang},
  \citenamefont {Zhao}, \citenamefont {Thiemens}, \citenamefont {Dominguez}
  \emph {et~al.}}]{fei2012gate}%
  \BibitemOpen
  \bibfield  {author} {\bibinfo {author} {\bibfnamefont {Z.}~\bibnamefont
  {Fei}}, \bibinfo {author} {\bibfnamefont {A.~S.}\ \bibnamefont {Rodin}},
  \bibinfo {author} {\bibfnamefont {G.~O.}\ \bibnamefont {Andreev}}, \bibinfo
  {author} {\bibfnamefont {W.}~\bibnamefont {Bao}}, \bibinfo {author}
  {\bibfnamefont {A.~S.}\ \bibnamefont {McLeod}}, \bibinfo {author}
  {\bibfnamefont {M.}~\bibnamefont {Wagner}}, \bibinfo {author} {\bibfnamefont
  {L.~M.}\ \bibnamefont {Zhang}}, \bibinfo {author} {\bibfnamefont
  {Z.}~\bibnamefont {Zhao}}, \bibinfo {author} {\bibfnamefont {M.}~\bibnamefont
  {Thiemens}}, \bibinfo {author} {\bibfnamefont {G.}~\bibnamefont {Dominguez}},
  \emph {et~al.},\ }\bibfield  {title} {\bibinfo {title} {Gate-tuning of
  graphene plasmons revealed by infrared nano-imaging},\ }\href@noop {}
  {\bibfield  {journal} {\bibinfo  {journal} {Nature}\ }\textbf {\bibinfo
  {volume} {487}},\ \bibinfo {pages} {82} (\bibinfo {year} {2012})}\BibitemShut
  {NoStop}%
\bibitem [{\citenamefont {Chen}\ \emph {et~al.}(2012)\citenamefont {Chen},
  \citenamefont {Badioli}, \citenamefont {Alonso-Gonz{\'a}lez}, \citenamefont
  {Thongrattanasiri}, \citenamefont {Huth}, \citenamefont {Osmond},
  \citenamefont {Spasenovi{\'c}}, \citenamefont {Centeno}, \citenamefont
  {Pesquera}, \citenamefont {Godignon} \emph {et~al.}}]{chen2012optical}%
  \BibitemOpen
  \bibfield  {author} {\bibinfo {author} {\bibfnamefont {J.}~\bibnamefont
  {Chen}}, \bibinfo {author} {\bibfnamefont {M.}~\bibnamefont {Badioli}},
  \bibinfo {author} {\bibfnamefont {P.}~\bibnamefont {Alonso-Gonz{\'a}lez}},
  \bibinfo {author} {\bibfnamefont {S.}~\bibnamefont {Thongrattanasiri}},
  \bibinfo {author} {\bibfnamefont {F.}~\bibnamefont {Huth}}, \bibinfo {author}
  {\bibfnamefont {J.}~\bibnamefont {Osmond}}, \bibinfo {author} {\bibfnamefont
  {M.}~\bibnamefont {Spasenovi{\'c}}}, \bibinfo {author} {\bibfnamefont
  {A.}~\bibnamefont {Centeno}}, \bibinfo {author} {\bibfnamefont
  {A.}~\bibnamefont {Pesquera}}, \bibinfo {author} {\bibfnamefont
  {P.}~\bibnamefont {Godignon}}, \emph {et~al.},\ }\bibfield  {title} {\bibinfo
  {title} {Optical nano-imaging of gate-tunable graphene plasmons},\
  }\href@noop {} {\bibfield  {journal} {\bibinfo  {journal} {Nature}\ }\textbf
  {\bibinfo {volume} {487}},\ \bibinfo {pages} {77} (\bibinfo {year}
  {2012})}\BibitemShut {NoStop}%
\bibitem [{\citenamefont {Woessner}\ \emph {et~al.}(2015)\citenamefont
  {Woessner}, \citenamefont {Lundeberg}, \citenamefont {Gao}, \citenamefont
  {Principi}, \citenamefont {Alonso-Gonz{\'a}lez}, \citenamefont {Carrega},
  \citenamefont {Watanabe}, \citenamefont {Taniguchi}, \citenamefont {Vignale},
  \citenamefont {Polini} \emph {et~al.}}]{woessner2015highly}%
  \BibitemOpen
  \bibfield  {author} {\bibinfo {author} {\bibfnamefont {A.}~\bibnamefont
  {Woessner}}, \bibinfo {author} {\bibfnamefont {M.~B.}\ \bibnamefont
  {Lundeberg}}, \bibinfo {author} {\bibfnamefont {Y.}~\bibnamefont {Gao}},
  \bibinfo {author} {\bibfnamefont {A.}~\bibnamefont {Principi}}, \bibinfo
  {author} {\bibfnamefont {P.}~\bibnamefont {Alonso-Gonz{\'a}lez}}, \bibinfo
  {author} {\bibfnamefont {M.}~\bibnamefont {Carrega}}, \bibinfo {author}
  {\bibfnamefont {K.}~\bibnamefont {Watanabe}}, \bibinfo {author}
  {\bibfnamefont {T.}~\bibnamefont {Taniguchi}}, \bibinfo {author}
  {\bibfnamefont {G.}~\bibnamefont {Vignale}}, \bibinfo {author} {\bibfnamefont
  {M.}~\bibnamefont {Polini}}, \emph {et~al.},\ }\bibfield  {title} {\bibinfo
  {title} {Highly confined low-loss plasmons in graphene--boron nitride
  heterostructures},\ }\href@noop {} {\bibfield  {journal} {\bibinfo  {journal}
  {Nature Materials}\ }\textbf {\bibinfo {volume} {14}},\ \bibinfo {pages}
  {421} (\bibinfo {year} {2015})}\BibitemShut {NoStop}%
\bibitem [{\citenamefont {Grigorenko}\ \emph {et~al.}(2012)\citenamefont
  {Grigorenko}, \citenamefont {Polini},\ and\ \citenamefont
  {Novoselov}}]{grigorenko2012graphene}%
  \BibitemOpen
  \bibfield  {author} {\bibinfo {author} {\bibfnamefont {A.~N.}\ \bibnamefont
  {Grigorenko}}, \bibinfo {author} {\bibfnamefont {M.}~\bibnamefont {Polini}},\
  and\ \bibinfo {author} {\bibfnamefont {K.~S.}\ \bibnamefont {Novoselov}},\
  }\bibfield  {title} {\bibinfo {title} {Graphene plasmonics},\ }\href@noop {}
  {\bibfield  {journal} {\bibinfo  {journal} {Nature Photonics}\ }\textbf
  {\bibinfo {volume} {6}},\ \bibinfo {pages} {749} (\bibinfo {year}
  {2012})}\BibitemShut {NoStop}%
\bibitem [{\citenamefont {Principi}\ \emph {et~al.}(2014)\citenamefont
  {Principi}, \citenamefont {Vignale}, \citenamefont {Carrega},\ and\
  \citenamefont {Polini}}]{principi2014plasmon}%
  \BibitemOpen
  \bibfield  {author} {\bibinfo {author} {\bibfnamefont {A.}~\bibnamefont
  {Principi}}, \bibinfo {author} {\bibfnamefont {G.}~\bibnamefont {Vignale}},
  \bibinfo {author} {\bibfnamefont {M.}~\bibnamefont {Carrega}},\ and\ \bibinfo
  {author} {\bibfnamefont {M.}~\bibnamefont {Polini}},\ }\bibfield  {title}
  {\bibinfo {title} {Plasmon losses due to electron--phonon scattering in
  graphene},\ }\href@noop {} {\bibfield  {journal} {\bibinfo  {journal}
  {Physical Review B}\ }\textbf {\bibinfo {volume} {90}},\ \bibinfo {pages}
  {165408} (\bibinfo {year} {2014})}\BibitemShut {NoStop}%
\bibitem [{\citenamefont {Yu}\ \emph {et~al.}(2008)\citenamefont {Yu},
  \citenamefont {Veronis}, \citenamefont {Fan},\ and\ \citenamefont
  {Brongersma}}]{yu2008gain}%
  \BibitemOpen
  \bibfield  {author} {\bibinfo {author} {\bibfnamefont {Z.}~\bibnamefont
  {Yu}}, \bibinfo {author} {\bibfnamefont {G.}~\bibnamefont {Veronis}},
  \bibinfo {author} {\bibfnamefont {S.}~\bibnamefont {Fan}},\ and\ \bibinfo
  {author} {\bibfnamefont {M.~L.}\ \bibnamefont {Brongersma}},\ }\bibfield
  {title} {\bibinfo {title} {Gain-induced switching in metal-dielectric-metal
  plasmonic waveguides},\ }\href@noop {} {\bibfield  {journal} {\bibinfo
  {journal} {Applied Physics Letters}\ }\textbf {\bibinfo {volume} {92}},\
  \bibinfo {pages} {041117} (\bibinfo {year} {2008})}\BibitemShut {NoStop}%
\bibitem [{\citenamefont {Athanasopoulos}\ \emph {et~al.}(2014)\citenamefont
  {Athanasopoulos}, \citenamefont {Mattheakis},\ and\ \citenamefont
  {Tsironis}}]{athanasopoulos2013enhanced}%
  \BibitemOpen
  \bibfield  {author} {\bibinfo {author} {\bibfnamefont {C.}~\bibnamefont
  {Athanasopoulos}}, \bibinfo {author} {\bibfnamefont {M.}~\bibnamefont
  {Mattheakis}},\ and\ \bibinfo {author} {\bibfnamefont {G.~P.}\ \bibnamefont
  {Tsironis}},\ }\bibfield  {title} {\bibinfo {title} {Surface plasmons on
  dielectric-metal-dielectric waveguides},\ }in\ \href
  {http://www.comsol.com/conference2014/europe/presentations/} {\emph {\bibinfo
  {booktitle} {Proceedings of the 2014 COMSOL Conference in Cambridge}}}\
  (\bibinfo {year} {2014})\ \bibinfo {note} {17--19 September;
  arXiv:1311.5722}\BibitemShut {NoStop}%
\bibitem [{\citenamefont {Bergman}\ and\ \citenamefont
  {Stockman}(2003)}]{bergman2003spaser}%
  \BibitemOpen
  \bibfield  {author} {\bibinfo {author} {\bibfnamefont {D.~J.}\ \bibnamefont
  {Bergman}}\ and\ \bibinfo {author} {\bibfnamefont {M.~I.}\ \bibnamefont
  {Stockman}},\ }\bibfield  {title} {\bibinfo {title} {Surface plasmon
  amplification by stimulated emission of radiation: quantum generation of
  coherent surface plasmons in nanosystems},\ }\href@noop {} {\bibfield
  {journal} {\bibinfo  {journal} {Physical Review Letters}\ }\textbf {\bibinfo
  {volume} {90}},\ \bibinfo {pages} {027402} (\bibinfo {year}
  {2003})}\BibitemShut {NoStop}%
\bibitem [{\citenamefont {Noginov}\ \emph {et~al.}(2009)\citenamefont
  {Noginov}, \citenamefont {Zhu}, \citenamefont {Belgrave}, \citenamefont
  {Bakker}, \citenamefont {Shalaev}, \citenamefont {Narimanov}, \citenamefont
  {Stout}, \citenamefont {Herz}, \citenamefont {Suteewong},\ and\ \citenamefont
  {Wiesner}}]{noginov2009spaser}%
  \BibitemOpen
  \bibfield  {author} {\bibinfo {author} {\bibfnamefont {M.~A.}\ \bibnamefont
  {Noginov}}, \bibinfo {author} {\bibfnamefont {G.}~\bibnamefont {Zhu}},
  \bibinfo {author} {\bibfnamefont {A.~M.}\ \bibnamefont {Belgrave}}, \bibinfo
  {author} {\bibfnamefont {R.}~\bibnamefont {Bakker}}, \bibinfo {author}
  {\bibfnamefont {V.~M.}\ \bibnamefont {Shalaev}}, \bibinfo {author}
  {\bibfnamefont {E.~E.}\ \bibnamefont {Narimanov}}, \bibinfo {author}
  {\bibfnamefont {S.}~\bibnamefont {Stout}}, \bibinfo {author} {\bibfnamefont
  {E.}~\bibnamefont {Herz}}, \bibinfo {author} {\bibfnamefont {T.}~\bibnamefont
  {Suteewong}},\ and\ \bibinfo {author} {\bibfnamefont {U.}~\bibnamefont
  {Wiesner}},\ }\bibfield  {title} {\bibinfo {title} {Demonstration of a
  spaser-based nanolaser},\ }\href@noop {} {\bibfield  {journal} {\bibinfo
  {journal} {Nature}\ }\textbf {\bibinfo {volume} {460}},\ \bibinfo {pages}
  {1110} (\bibinfo {year} {2009})}\BibitemShut {NoStop}%
\bibitem [{\citenamefont {Oulton}\ \emph {et~al.}(2009)\citenamefont {Oulton},
  \citenamefont {Sorger}, \citenamefont {Zentgraf}, \citenamefont {Ma},
  \citenamefont {Gladden}, \citenamefont {Dai}, \citenamefont {Bartal},\ and\
  \citenamefont {Zhang}}]{oulton2009plasmon}%
  \BibitemOpen
  \bibfield  {author} {\bibinfo {author} {\bibfnamefont {R.~F.}\ \bibnamefont
  {Oulton}}, \bibinfo {author} {\bibfnamefont {V.~J.}\ \bibnamefont {Sorger}},
  \bibinfo {author} {\bibfnamefont {T.}~\bibnamefont {Zentgraf}}, \bibinfo
  {author} {\bibfnamefont {R.-M.}\ \bibnamefont {Ma}}, \bibinfo {author}
  {\bibfnamefont {C.}~\bibnamefont {Gladden}}, \bibinfo {author} {\bibfnamefont
  {L.}~\bibnamefont {Dai}}, \bibinfo {author} {\bibfnamefont {G.}~\bibnamefont
  {Bartal}},\ and\ \bibinfo {author} {\bibfnamefont {X.}~\bibnamefont
  {Zhang}},\ }\bibfield  {title} {\bibinfo {title} {Plasmon lasers at deep
  subwavelength scale},\ }\href {https://doi.org/10.1038/nature08364}
  {\bibfield  {journal} {\bibinfo  {journal} {Nature}\ }\textbf {\bibinfo
  {volume} {461}},\ \bibinfo {pages} {629} (\bibinfo {year}
  {2009})}\BibitemShut {NoStop}%
\bibitem [{\citenamefont {Menendez}\ \emph {et~al.}(2016)\citenamefont
  {Menendez}, \citenamefont {Rosolen},\ and\ \citenamefont
  {Maes}}]{menendez2016graphene}%
  \BibitemOpen
  \bibfield  {author} {\bibinfo {author} {\bibfnamefont {G.~A.}\ \bibnamefont
  {Menendez}}, \bibinfo {author} {\bibfnamefont {G.}~\bibnamefont {Rosolen}},\
  and\ \bibinfo {author} {\bibfnamefont {B.}~\bibnamefont {Maes}},\ }\bibfield
  {title} {\bibinfo {title} {Graphene plasmons embedded in a gain medium: layer
  and ribbon plasmons},\ }\href@noop {} {\bibfield  {journal} {\bibinfo
  {journal} {Journal of Optics}\ }\textbf {\bibinfo {volume} {18}},\ \bibinfo
  {pages} {125004} (\bibinfo {year} {2016})}\BibitemShut {NoStop}%
\bibitem [{\citenamefont {Zhernovnykova}\ \emph {et~al.}(2019)\citenamefont
  {Zhernovnykova}, \citenamefont {Popova}, \citenamefont {Deynychenko},
  \citenamefont {Deynichenko},\ and\ \citenamefont
  {Bludov}}]{zhernovnykova2019surface}%
  \BibitemOpen
  \bibfield  {author} {\bibinfo {author} {\bibfnamefont {O.}~\bibnamefont
  {Zhernovnykova}}, \bibinfo {author} {\bibfnamefont {O.}~\bibnamefont
  {Popova}}, \bibinfo {author} {\bibfnamefont {G.}~\bibnamefont {Deynychenko}},
  \bibinfo {author} {\bibfnamefont {T.}~\bibnamefont {Deynichenko}},\ and\
  \bibinfo {author} {\bibfnamefont {Y.~V.}\ \bibnamefont {Bludov}},\ }\bibfield
   {title} {\bibinfo {title} {Surface plasmon-polaritons in graphene, embedded
  into medium with gain and losses},\ }\href@noop {} {\bibfield  {journal}
  {\bibinfo  {journal} {Journal of Physics: Condensed Matter}\ }\textbf
  {\bibinfo {volume} {31}},\ \bibinfo {pages} {465301} (\bibinfo {year}
  {2019})}\BibitemShut {NoStop}%
\bibitem [{\citenamefont {Mattheakis}\ \emph {et~al.}(2017)\citenamefont
  {Mattheakis}, \citenamefont {Tsironis},\ and\ \citenamefont
  {Kaxiras}}]{mattheakis2017chapter}%
  \BibitemOpen
  \bibfield  {author} {\bibinfo {author} {\bibfnamefont {M.}~\bibnamefont
  {Mattheakis}}, \bibinfo {author} {\bibfnamefont {G.~P.}\ \bibnamefont
  {Tsironis}},\ and\ \bibinfo {author} {\bibfnamefont {E.}~\bibnamefont
  {Kaxiras}},\ }\bibfield  {title} {\bibinfo {title} {Graphene and active
  metamaterials: Theoretical methods and physical properties},\ }in\ \href
  {https://doi.org/10.5772/67900} {\emph {\bibinfo {booktitle} {Nanoplasmonics:
  Fundamentals and Applications}}},\ \bibinfo {editor} {edited by\ \bibinfo
  {editor} {\bibfnamefont {G.}~\bibnamefont {Barbillon}}}\ (\bibinfo
  {publisher} {IntechOpen},\ \bibinfo {year} {2017})\BibitemShut {NoStop}%
\bibitem [{\citenamefont {Falkovsky}(2008)}]{falkovsky2008optical}%
  \BibitemOpen
  \bibfield  {author} {\bibinfo {author} {\bibfnamefont {L.}~\bibnamefont
  {Falkovsky}},\ }\bibfield  {title} {\bibinfo {title} {Optical properties of
  doped graphene layers},\ }\href@noop {} {\bibfield  {journal} {\bibinfo
  {journal} {Journal of Experimental and Theoretical Physics}\ }\textbf
  {\bibinfo {volume} {106}},\ \bibinfo {pages} {575} (\bibinfo {year}
  {2008})}\BibitemShut {NoStop}%
\bibitem [{\citenamefont {Gusynin}\ \emph {et~al.}(2007)\citenamefont
  {Gusynin}, \citenamefont {Sharapov},\ and\ \citenamefont
  {Carbotte}}]{gusynin2007ac}%
  \BibitemOpen
  \bibfield  {author} {\bibinfo {author} {\bibfnamefont {V.~P.}\ \bibnamefont
  {Gusynin}}, \bibinfo {author} {\bibfnamefont {S.~G.}\ \bibnamefont
  {Sharapov}},\ and\ \bibinfo {author} {\bibfnamefont {J.~P.}\ \bibnamefont
  {Carbotte}},\ }\bibfield  {title} {\bibinfo {title} {Ac conductivity of
  graphene: from tight-binding model to 2+1-dimensional quantum
  electrodynamics},\ }\href@noop {} {\bibfield  {journal} {\bibinfo  {journal}
  {International Journal of Modern Physics B}\ }\textbf {\bibinfo {volume}
  {21}},\ \bibinfo {pages} {4611} (\bibinfo {year} {2007})}\BibitemShut
  {NoStop}%
\bibitem [{\citenamefont {Hanson}(2008)}]{hanson2008dyadic}%
  \BibitemOpen
  \bibfield  {author} {\bibinfo {author} {\bibfnamefont {G.~W.}\ \bibnamefont
  {Hanson}},\ }\bibfield  {title} {\bibinfo {title} {Dyadic green's functions
  and guided surface waves for a surface conductivity model of graphene},\
  }\href@noop {} {\bibfield  {journal} {\bibinfo  {journal} {Journal of Applied
  Physics}\ }\textbf {\bibinfo {volume} {103}},\ \bibinfo {pages} {064302}
  (\bibinfo {year} {2008})}\BibitemShut {NoStop}%
\bibitem [{\citenamefont {Profumo}\ \emph {et~al.}(2012)\citenamefont
  {Profumo}, \citenamefont {Asgari}, \citenamefont {Polini},\ and\
  \citenamefont {MacDonald}}]{profumo2012double}%
  \BibitemOpen
  \bibfield  {author} {\bibinfo {author} {\bibfnamefont {R.~E.~V.}\
  \bibnamefont {Profumo}}, \bibinfo {author} {\bibfnamefont {R.}~\bibnamefont
  {Asgari}}, \bibinfo {author} {\bibfnamefont {M.}~\bibnamefont {Polini}},\
  and\ \bibinfo {author} {\bibfnamefont {A.~H.}\ \bibnamefont {MacDonald}},\
  }\bibfield  {title} {\bibinfo {title} {Double-layer graphene and topological
  insulator thin-film plasmons},\ }\href
  {https://doi.org/10.1103/PhysRevB.85.085443} {\bibfield  {journal} {\bibinfo
  {journal} {Physical Review B}\ }\textbf {\bibinfo {volume} {85}},\ \bibinfo
  {pages} {085443} (\bibinfo {year} {2012})}\BibitemShut {NoStop}%
\bibitem [{\citenamefont {Hwang}\ and\ \citenamefont
  {Das~Sarma}(2007)}]{hwang2007dielectric}%
  \BibitemOpen
  \bibfield  {author} {\bibinfo {author} {\bibfnamefont {E.~H.}\ \bibnamefont
  {Hwang}}\ and\ \bibinfo {author} {\bibfnamefont {S.}~\bibnamefont
  {Das~Sarma}},\ }\bibfield  {title} {\bibinfo {title} {Dielectric function,
  screening, and plasmons in two-dimensional graphene},\ }\href
  {https://doi.org/10.1103/PhysRevB.75.205418} {\bibfield  {journal} {\bibinfo
  {journal} {Physical Review B}\ }\textbf {\bibinfo {volume} {75}},\ \bibinfo
  {pages} {205418} (\bibinfo {year} {2007})}\BibitemShut {NoStop}%
\end{thebibliography}
\end{document}